\documentclass[prb,twocolumn,showpacs,amsmath,amssymb,superscriptaddress,floatfix]{revtex4-1}

\usepackage[dvipdfmx]{graphicx}
\usepackage{graphicx}
\usepackage{dcolumn}
\usepackage{bm}
\usepackage{epsf}
\usepackage{color}
\usepackage{subfigure}

\begin{document}

\preprint{preprint}

\title{Hybridized and Localized 4\textbf{\textit{f}} Electronic States of Nd-based Intermetallic Compounds in Cubic Symmetry Probed by High-Energy Photoemission}

\author{M. Sakaguchi}
\affiliation{Division of Materials Physics, Graduate School of Engineering Science, The University of Osaka, Toyonaka, Osaka 560-8531, Japan}
\affiliation{RIKEN SPring-8 Center, Sayo, Hyogo 679-5148, Japan}
\author{A. Enomoto}
\affiliation{Division of Materials Physics, Graduate School of Engineering Science, The University of Osaka, Toyonaka, Osaka 560-8531, Japan}
\affiliation{RIKEN SPring-8 Center, Sayo, Hyogo 679-5148, Japan}
\author{H. Fujiwara}
\affiliation{Division of Materials Physics, Graduate School of Engineering Science, The University of Osaka, Toyonaka, Osaka 560-8531, Japan}
\affiliation{RIKEN SPring-8 Center, Sayo, Hyogo 679-5148, Japan}
\affiliation{Spintronics Research Network Division, Institute for Open and Transdisciplinary Research Initiatives, The University of Osaka, Suita, Osaka 565-0871, Japan}
\author{G. Nozue}
\affiliation{Division of Materials Physics, Graduate School of Engineering Science, The University of Osaka, Toyonaka, Osaka 560-8531, Japan}
\affiliation{RIKEN SPring-8 Center, Sayo, Hyogo 679-5148, Japan}
\author{S. Hamamoto}
\affiliation{RIKEN SPring-8 Center, Sayo, Hyogo 679-5148, Japan}
\author{Y. Torii}
\affiliation{Division of Materials Physics, Graduate School of Engineering Science, The University of Osaka, Toyonaka, Osaka 560-8531, Japan}
\affiliation{RIKEN SPring-8 Center, Sayo, Hyogo 679-5148, Japan}
\author{T. D. Nakamura}
\affiliation{Division of Materials Physics, Graduate School of Engineering Science, The University of Osaka, Toyonaka, Osaka 560-8531, Japan}
\affiliation{RIKEN SPring-8 Center, Sayo, Hyogo 679-5148, Japan}
\author{N. U. Sakamoto}
\affiliation{Division of Materials Physics, Graduate School of Engineering Science, The University of Osaka, Toyonaka, Osaka 560-8531, Japan}
\affiliation{RIKEN SPring-8 Center, Sayo, Hyogo 679-5148, Japan}
\author{K. Yamagami}
\affiliation{Division of Materials Physics, Graduate School of Engineering Science, The University of Osaka, Toyonaka, Osaka 560-8531, Japan}
\affiliation{Japan Synchrotron Radiation Research Institute, Sayo, Hyogo 679-5198, Japan}
\author{T. Kiss}
\affiliation{Division of Materials Physics, Graduate School of Engineering Science, The University of Osaka, Toyonaka, Osaka 560-8531, Japan}
\author{Y. Kanai-Nakata}
\affiliation{RIKEN SPring-8 Center, Sayo, Hyogo 679-5148, Japan}
\affiliation{Department of Physical Sciences, Faculty of Science and Engineering, Ritsumeikan University, Kusatsu, Shiga 525-8577, Japan}
\author{S. Imada}
\affiliation{RIKEN SPring-8 Center, Sayo, Hyogo 679-5148, Japan}
\affiliation{Department of Physical Sciences, Faculty of Science and Engineering, Ritsumeikan University, Kusatsu, Shiga 525-8577, Japan}
\author{A. Irizawa}
\affiliation{Research Organization of Science and Technology, Ritsumeikan University, Kusatsu, Shiga 525-8577, Japan}
\author{A. Yamasaki}
\affiliation{RIKEN SPring-8 Center, Sayo, Hyogo 679-5148, Japan}
\affiliation{Faculty of Science and Engineering, Konan University, Kobe, Hyogo 658-8501, Japan}
\author{A. Higashiya}
\affiliation{RIKEN SPring-8 Center, Sayo, Hyogo 679-5148, Japan}
\affiliation{Faculty of Science and Engineering, Setsunan University, Neyagawa, Osaka 572-8508, Japan}
\author{M. Oura}
\affiliation{RIKEN SPring-8 Center, Sayo, Hyogo 679-5148, Japan}
\author{K. Tamasaku}
\affiliation{RIKEN SPring-8 Center, Sayo, Hyogo 679-5148, Japan}
\author{M. Yabashi}
\affiliation{RIKEN SPring-8 Center, Sayo, Hyogo 679-5148, Japan}
\author{T. Ishikawa}
\affiliation{RIKEN SPring-8 Center, Sayo, Hyogo 679-5148, Japan}
\author{H. Sugawara}
\affiliation{Department of Physics, Kobe University, Kobe, Hyogo 657-8501, Japan}
\author{H. Amitsuka}
\affiliation{Department of Physics, Hokkaido University, Sapporo, Hokkaido 060-0810, Japan}
\author{T. Yanagisawa}
\affiliation{Department of Physics, Hokkaido University, Sapporo, Hokkaido 060-0810, Japan}
\author{H. Hidaka}
\affiliation{Department of Physics, Hokkaido University, Sapporo, Hokkaido 060-0810, Japan}
\author{A. Sekiyama} 
\affiliation{Division of Materials Physics, Graduate School of Engineering Science, The University of Osaka, Toyonaka, Osaka 560-8531, Japan}
\affiliation{RIKEN SPring-8 Center, Sayo, Hyogo 679-5148, Japan}
\affiliation{Spintronics Research Network Division, Institute for Open and Transdisciplinary Research Initiatives, The University of Osaka, Suita, Osaka 565-0871, Japan}

\date{\today}

\begin{abstract}
We have performed soft and hard X-ray photoemission spectroscopies on NdTi$_2$Al$_{20}$ and NdBe$_{13}$ which show antiferromagnetic ordering at low temperatures. 
The Nd $3d$ core-level photoemission and Nd $3d$-$4f$ valence-band resonant photoemission spectra show finite $4f^4$ initial-state components in addition to the $4f^3$ configurations attributed to the $c$-$f$ hybridization effects in NdTi$_2$Al$_{20}$, while the $4f^3$ initial-state components with localized character are dominant in NdBe$_{13}$. 
These results imply the emergence of overscreening channel due to the two-channel Kondo effect in NdTi$_2$Al$_{20}$ through the the strong $c$-$f$ hybridization effect.
\end{abstract}





\maketitle

\section{INTRODUCTION}
In rare-earth-based strongly correlated electron systems with partially filled 4$f$ orbitals, 4$f$ electrons play a crucial role for the emergence of various physical properties. 
Since the 4$f$ electrons are spatially distributed relatively closer to the rare-earth nucleus than $5p$ inner-core electrons, they exhibit localized characteristics. 
However, at low temperatures, their itinerant nature becomes more pronounced through hybridization between the $4f$ and conduction electrons at the Fermi level ($E_F$).  
(Hereafter it is called as the $c$-$f$ hybridization.) 
Various phenomena induced by the $c$-$f$ hybridization have been observed. 
One of such phenomena is the heavy Fermionic nature due to the Kondo effect, in which the localized magnetic moments of  $4f$ electrons in rare-earth ions are screened by the surrounding conduction electrons through the $c$-$f$ hybridization.\cite{Kondo1,Kondo2} 
This anomaly is primary discussed in the simple cases, such as Ce and Yb compounds with one electron/hole in the $4f$ shell, respectively.\cite{Kondo_Yb,Kondo_Ce} 
In contrast, Nd compounds, which usually contain three 4$f$ electrons in the trivalent states, have recently been focused on the exotic phenomena such as cooperation of the magnetic ordering with non-Fermi liquid behavior, which arises from the overscreening of the localized moment by the conduction electrons.\cite{NdCo} 
This anomalous screening effect is theoretically predicted as so called two-channel Kondo effect, in which the quadrupole degree of freedom, treated as a quasi-spin, hybridizes with two types of conduction electrons in particular symmetry of the crystalline-electric-field (CEF)-split ground states.\cite{Two_Cox}
 
In the Nd compounds under cubic symmetry, the multiplets of the Nd ions with the total angular momentum $J=9/2$ split into three states, $\Gamma_6$, $\Gamma_8^{(1)}$, and $\Gamma_8^{(2)}$, due to the CEF.\cite{LLW}  
According to the previous study, the Nd compounds with the $\Gamma_6$ ground state exhibit a residual entropy of $\ln\sqrt{2}$ which is a characteristic of the two-channel Kondo effect.\cite{Kondo_Nd} 
Therefore, to experimentally investigate the possibility of the two-channel Kondo effect, it is important to identify the CEF-split ground state. However, this assumption is based on the premise that the $c$-$f$ hybridization works. Thus, it is also important to investigate the presence of $c$-$f$ hybridization. 

NdTi$_2$Al$_{20}$ is recognized as one of the candidate materials for the two-channel Kondo effect from the perspective of its crystal structure and the proposed CEF ground state.\cite{Nd1220_crystal,Nd1220_Sugawara2020}  
The crystal structure of NdTi$_2$Al$_{20}$ is cubic CeCr$_2$Al$_{20}$-type with the space group $Fd\bar{3}m$, where the Nd sites are under the $T_d$ symmetry. 
NdTi$_2$Al$_{20}$ exhibits an antiferromagnetic (AFM) order at $T_N = 1.45 $K.\cite{Nd1220_crystal} 
The magnetic part of the entropy $S_m(T)$ for NdTi$_2$Al$_{20}$ quickly increases with increasing temperature and reaches  $R\ln2$ at $T_N$, suggesting the $\Gamma_6$ doublet to be the CEF-split ground state.\cite{Nd1220_crystal} 
In contrast, NdBe$_{13}$, which is considered to have weak $c$-$f$ hybridization, 
crystallizes in the cubic NdZn$_{13}$ structure with the space group $Fm\bar{3}c$ in which the Nd sites are in the $O$ symmetry.\cite{NaZn13_cry,NdBe13_crystal} 
This compound undergoes an AFM ordering at $T_N$ = 2.6 K,\cite{NdBe13_AFM} of which the magnetization measurements have suggested the $\Gamma_8^{(2)}$ symmetry.\cite{NdBe13_crystal}

In this paper, we have performed Nd $3d$ core-level soft and hard X-ray photoemission spectroscopy (SXPES, HAXPES) for NdTi$_2$Al$_{20}$ and NdBe$_{13}$ to selectively investigate the local electronic structures of the Nd $4f$ states. 
We have also performed X-ray absorption spectroscopy (XAS) and valence-band resonant photoemission spectroscopy (RPES) to clarify the Nd 4$f$ state.

\section{Experimental}
Single crystals of NdTi$_2$Al$_{20}$ and NdBe$_{13}$ were synthesized by the Al-flux methods.\cite{Nd1220_Sugawara2020,HidakaUnpub}  
The Nd $M_{45}$-edge XAS and Nd 3$d$-4$f$ RPES were performed at BL17SU in SPring-8.\cite{BL17_1,BL17_2,BL17_3}  
The XAS spectra were obtained in the total-electron-yield mode. 
The RPES spectra were recorded by a SCIENTA EW4000 with the overall energy resolution of 170 meV. 
The measuring temperatures were set to 80 K and 11 K for NdTi$_2$Al$_{20}$ and NdBe$_{13}$, respectively. The clean surfaces were obtained by $in$-$situ$ fracturing at the base pressure of $1.3\times10^{-8}$ Pa.
The HAXPES experiments were performed at BL19LXU of SPring-8 using a MBS A1-HE hemispherical photoelectron spectrometer.\cite{BL19} 
The energy resolution was set to $400-500$ meV with the measuring temperature of $\sim$5 K. 
The clean surface was obtained by $in$-$situ$ fracturing at the base pressure of better than $1.2\times10^{-7}$ Pa.
We conformed that neither O 1$s$ nor C 1$s$ photoemission spectral weight due to the surface oxidization or the impurity in the alloy was negligible. 
We also measured the Nd $3d$ core-level HAXPES spectra of NdFe$_4$P$_{12}$ and Nd$_{1.85}$Ce$_{0.15}$CuO$_4$ at BL19LXU in SPring-8 for mutual comparison. 
In addition, the Nd $3d$ core-level HAXPES spectrum of NdGaO$_3$ was obtained at SA-1 in SR center of Ritsumeikan University.\cite{Ritsu}  

\section{RESULTS AND DISCUSSIONS}
\subsection{Core-level photoemission}
\begin{figure}[t] 
  \centering
       \includegraphics[keepaspectratio,width=80mm]{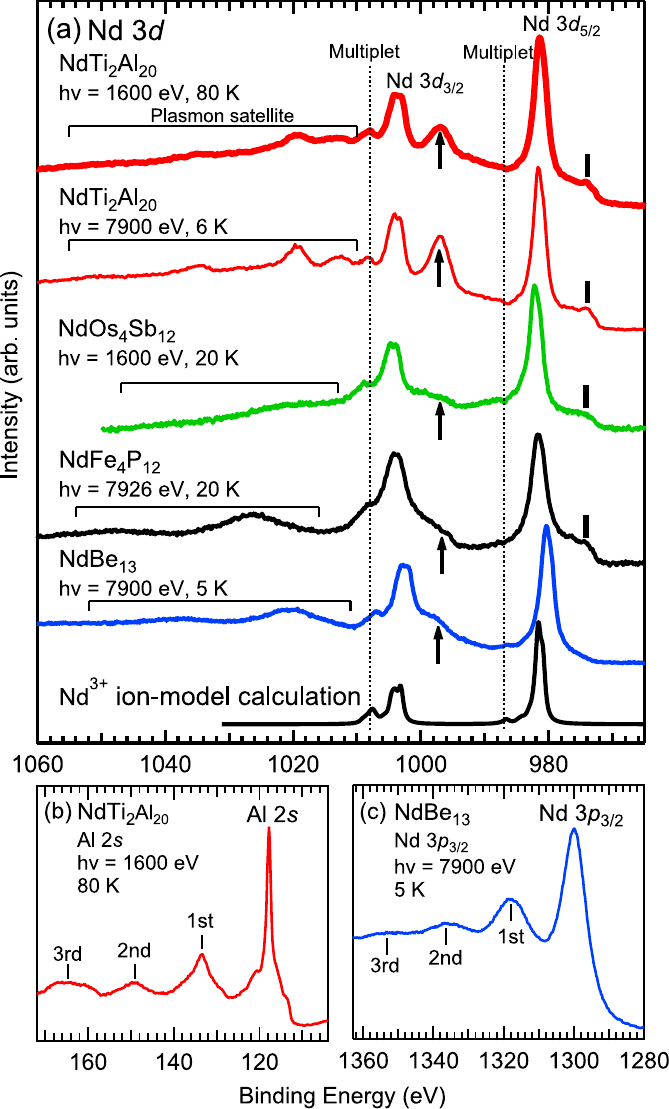}
 \caption
 {(a)Nd 3$d$ core-level SXPES spectrum of NdTi$_2$Al$_{20}$ compared with that of NdOs$_4$Sb$_{12}$ (Ref.\onlinecite{NdOsSb}), and HAXPES spectra of NdTi$_2$Al$_{20}$, NdFe$_4$P$_{12}$, and NdBe$_{13}$ together with the calculated Nd$^{3+}$ 3$d$ PES spectrum assuming the $3d^{10}4f^{3}\to3d^9 4f^3$ process. 
The black dashed lines indicate the multiplet components split from the main peaks.  
(b)Al 2$s$ core-level SXPES spectrum of NdTi$_2$Al$_{20}$. 
(c)Nd 3$p_{3/2}$ core-level HAXPES spectrum of NdBe$_{13}$.} 
   \label{Nd3dw}
\end{figure}
\begin{figure}[t] 
  \centering
       \includegraphics[keepaspectratio,width=80mm]{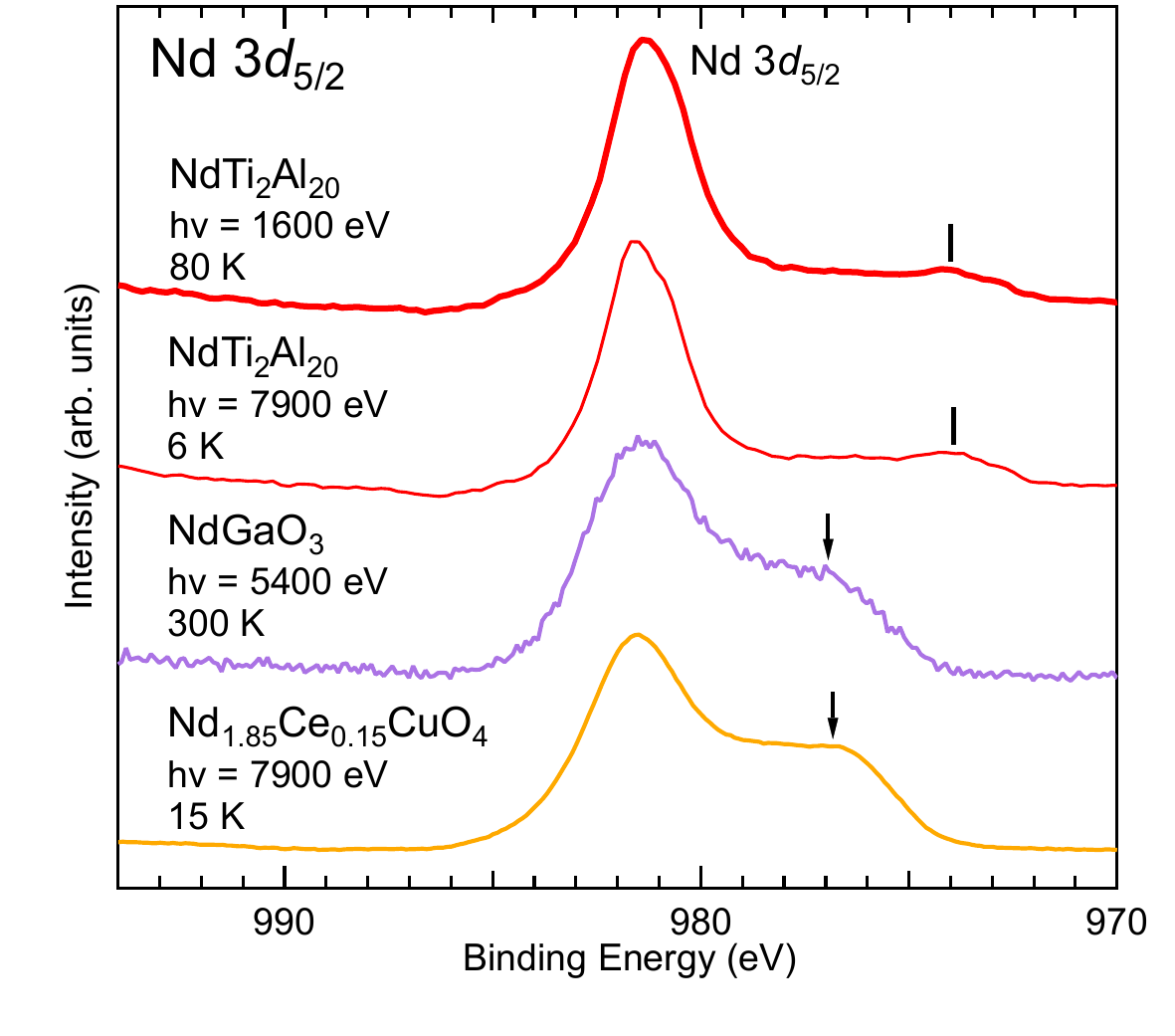}
 \caption
 {Nd $3d_{5/2}$ core-level PES spectra of NdTi$_2$Al$_{20}$, NdGaO$_3$, and Nd$_{1.85}$Ce$_{0.15}$CuO$_4$ after subtracting the Shirley-type backgrounds. The black line corresponds to the one in Fig. 1(a).  The downward arrow indicates the $3d^94f^4\underline{L}$ components originating from the charge transfer from the ligand ($L$) oxygen sites.} 
   \label{Nd3dn}
\end{figure}

Figure~\ref{Nd3dw}(a) shows the Nd 3$d$ core-level photoemission spectra of NdTi$_2$Al$_{20}$ and  NdBe$_{13}$. 
For both compounds, main peaks are seen at $\sim$980 and $\sim$1003 eV, corresponding to the Nd $3d_{5/2}$ and Nd $3d_{3/2}$ excitations, respectively. 
Weak shoulder structures are seen at $\sim$987 and $\sim$1008 eV, which are attributed to the multiplet components split from the main peaks. Indeed, these multiplet components are reproduced by the simulation of the ionic Nd$^{3+}$ $3d$ core-level photoemission explained below. 
Additionally, a broad hump structure is identified in the energy range of $1010-1055$ eV for both compounds. 
As shown in Figs.~\ref{Nd3dw}(b) and (c), similar hump features are also observed in the Al 2$s$ core-level SXPES spectrum of NdTi$_2$Al$_{20}$, and in the Nd 3$p_{3/2}$ core-level HAXPES spectrum of NdBe$_{13}$, appearing at energy shifts of 15.6 eV and 17.8 eV above the respective main peaks. 
The broad hump structures between $1010-1055$ eV are attributed to plasmon satellites associated with the Nd $3d_{5/2}$ and $3d_{3/2}$ peaks.
Moreover, a distinct spectral feature around 974 eV, indicated by the black line in Fig.~\ref{Nd3dw}(a), is observed in both the SXPES and HAXPES spectra of NdTi$_2$Al$_{20}$. 
Notably, this feature, in contrast to the Nd $3d_{5/2}$ main peak, cannot be explained by the theoretical Nd$^{3+}$ core-level PES spectrum, i.e., $3d^{10}4f^3\to3d^94f^3$ process obtained by the atomic multiplet calculation shown in the lower part of Fig.~\ref{Nd3dw}(a). 
On the other hand, this feature is also observed in the Nd-based filled skutterudite compounds NdOs$_4$Sb$_{12}$ (Ref.\onlinecite{NdOsSb}) and NdFe$_4$P$_{12}$, where the electronic specific heat coefficient $\gamma$ are 520 and 10 mJ/(mol$\cdot$K$^2$), respectively.\cite{G_NdOsSb,G_NdFeP}  
A structure at 997 eV, indicated by the upward arrow, is located at the same energy interval from the Nd $3d_{3/2}$ main peak as the interval between the shoulder structure at 974 eV and the Nd $3d_{5/2}$ main peak. 
The structure at 974 and 997 eV may have the same origin, but the first plasmon satellite of the Nd $3d_{5/2}$ main peaks appear also around 997 eV. 
That means,  it is rather hard to clearly discuss the origin of the structure at 997 eV, and thus we focus on the Nd $3d_{5/2}$ core-level spectra. 
Meanwhile, the Nd 3d$_{3/2}$ and 3d$_{5/2}$ main peaks of the spectra of NdBe$_{13}$, for which the shoulder at 974 eV is not recognized, are seen at the relatively shallower binding energy than those for NdTi$_2$Al$_{20}$, NdOs$_4$Sb$_{12}$, and NdFe$_4$P$_{12}$. 
This is discussed later together with the peaks in the valence-band RPES spectra. 
It should be noted that the similar tendency has been observed in the Pr $3d_{5/2}$ core-level main peaks between PrIr$_2$Zn$_{20}$ and PrBe$_{13}$.\cite{PrLD2} 

Figure~\ref{Nd3dn} shows the Nd $3d_{5/2}$ core-level PES spectra of NdTi$_2$Al$_{20}$. 
The spectral line shapes are mutually similar between at $h\nu =$ 1600 eV and at 7900 eV irrespective of much difference in the photoelectron kinetic energy, 
from which we conclude that the clearly observed shoulder at 974 eV is intrinsically reflecting the bulk Nd $4f$ electronic state. 
For comparison, the spectra of Nd oxides NdGaO$_3$ and Nd$_{1.85}$Ce$_{0.15}$CuO$_4$ are also plotted after subtracting the Shirley-type backgrounds in the figure.\cite{Ritsu} 
In the Nd oxides, shoulder structures are observed at 977 eV (indicated by a downward arrow in Fig.~\ref{Nd3dn}) in addition to the Nd $3d_{5/2}$ main peak at 981.5 eV. 
These structures are due to $3d^94f^4\underline{L}$ components originating from the charge transfer from the ligand ($L$) oxygen sites as as already discussed for the other Nd-based oxides.\cite{Ritsu,Suzuki,Horio} 
Therefore, the structures around 974 eV in NdTi$_2$Al$_{20}$  are not the oxidation-derived but intrinsic features of the $3d^94f^4\underline{c}$ configurations originating from the $c$-$f$ hybridization. 
The difference in the energy splitting of the shoulders from the main peaks between NdTi$_2$Al$_{20}$ and Nd-based oxides reflects the different energy levels of the conduction electron (nearly $E_F$) in NdTi$_2$Al$_{20}$ and the O $2p$ orbitals (well below $E_F$) in the Nd oxides as verified by the Nd $3d-4f$ RPES of another Nd oxide\cite{ASNSMO99} whereas the bare Nd$^{3+}$ $4f$ levels (before the $c$-$f$ hybridizations are switching on) are more or less well below $E_F$.

\subsection{Resonance photoemission}

\begin{figure}[t] 
  \centering
   \includegraphics[keepaspectratio,width=80mm]{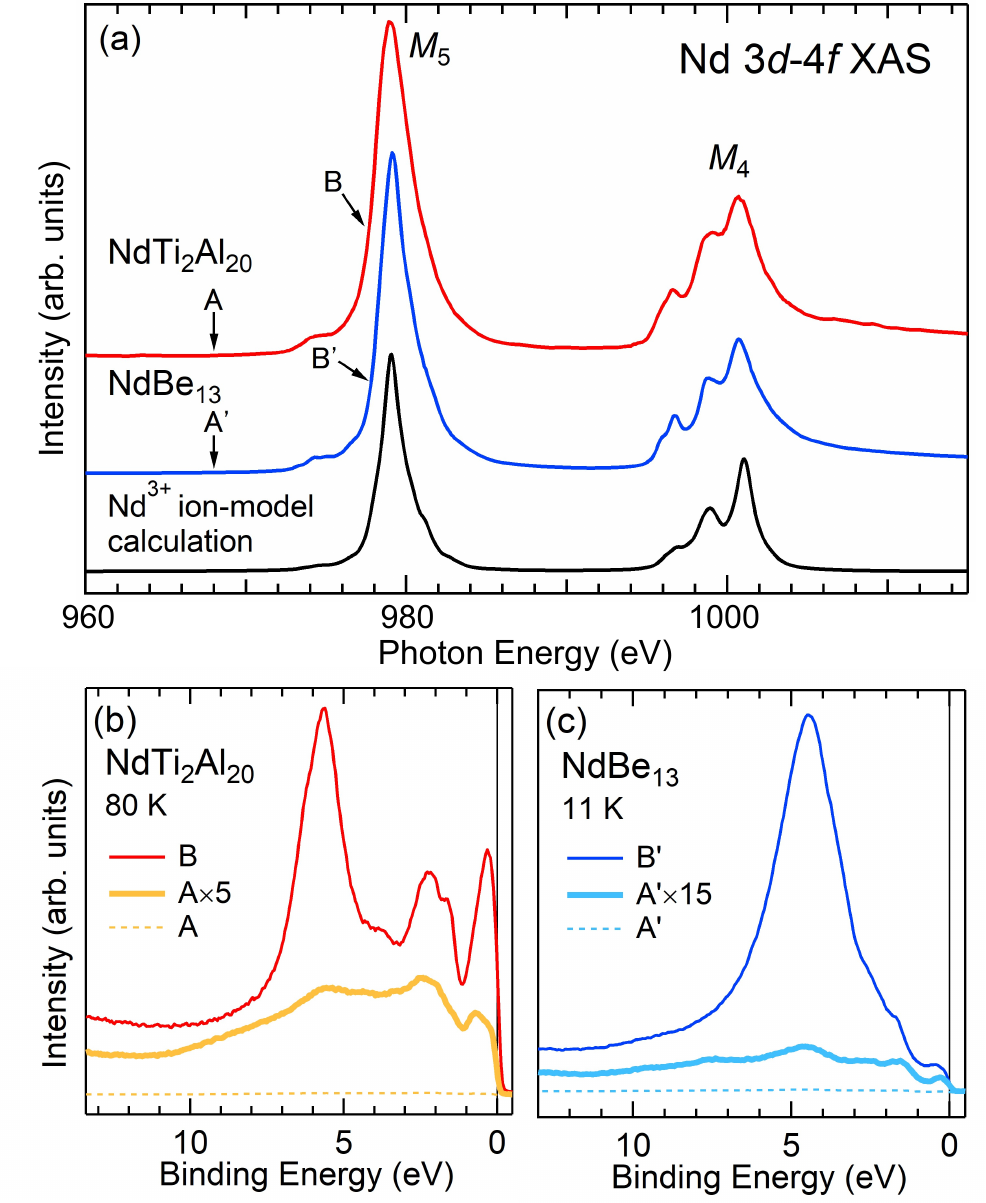}
 \caption
 {Nd $3d\to4f$ XAS spectra of NdTi$_2$Al$_{20}$ and NdBe$_{13}$ compared with the calculated Nd$^{3+}$ XAS assuming the $3d^{10}4f^{3}\to3d^94f^4$ process. The arrows with labels A, B, A$'$ and B$'$ indicate the selected photon energies at which the on- and off-RPES spectra of (b)NdTi$_2$Al$_{20}$ and (c)NdBe$_{13}$ have been measured.} 
   \label{XASandRPES}
\end{figure}

Figure~\ref{XASandRPES}(a) shows the experimental Nd $3d$-$4f$ XAS spectra of NdTi$_2$Al$_{20}$ and NdBe$_{13}$ together with the simulated spectrum by the Nd$^{3+}$ ion-model calculation. 
The calculation corresponds to the atomic multiplet model for the $3d^{10}4f^{3}\to3d^94f^4$ transition, incorporating both spin-orbit coupling and electron-electron interactions. 
The experimental XAS spectra for both compounds are fairly consistent with the ionic model simulation, suggesting that the Nd ions in these compounds are predominantly trivalent. 
In order to further study the contributions of the electrons in the Nd sites to the valence bands, we have mutually compared the on- and off-RPES spectra of NdTi$_2$Al$_{20}$ and  NdBe$_{13}$ in Figs.~\ref{XASandRPES}(b) and (c). 
The off (on)-RPES spectra of NdTi$_2$Al$_{20}$ and NdBe$_{13}$ are taken at photon energy labeled as A (B) and A$'$ (B$'$) in Fig.~\ref{XASandRPES}(a), respectively. 
For NdTi$_2$Al$_{20}$, the structure of on-RPES spectrum exhibits the resonance enhancement around $E_F$. 
Moreover, the structures around 6 eV as well as those near $E_F$ exhibit significant enhancement with changing the spectral lineshape due to the resonance effect. 
In contrast, for NdBe$_{13}$, only the structure around 5 eV shows a notable change whereas the spectral enhancements are relatively much smaller near $E_F$, reflecting the localized character of the Nd 4$f$ states in NdBe$_{13}$. 

\begin{figure}[t] 
  \centering
       \includegraphics[keepaspectratio,width=80mm]{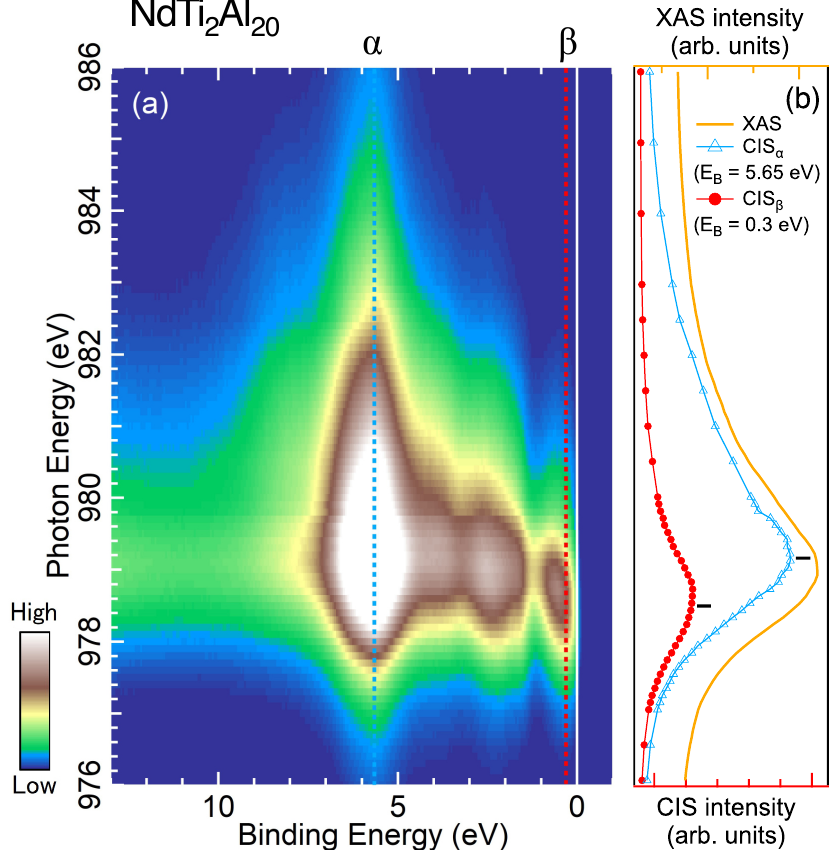}
 \caption
 {(a) Intensity plot of the valence-band RPES spectra of NdTi$_2$Al$_{20}$ taken with photon energies ranging from 968 to 990 eV across the Nd $M_5$ absorption edge. The spectral weight has been linearly interpolated with respect to the photon energy. (b)CIS spectra at $E_B$ = 5.65 ($\alpha$) and 0.3 eV ($\beta$). For comparison, the experimental XAS spectrum (yellow) is also plotted. The black lines indicate the peak photon energy of the CIS spectra.} 
   \label{RPES2Dmap1220}
\end{figure}

To discuss the origin of the resonance enhancement for NdTi$_2$Al$_{20}$, we have also performed detailed valence-band RPES measurements by finely tuning the incident photon energy $h\nu$. 
Figure~\ref{RPES2Dmap1220}(a) shows the intensity plot of RPES spectra of NdTi$_2$Al$_{20}$ across the $M_5$ absorption edge, with the incident photon energies ranging from 968 to 990 eV. 
The spectra are normalized by the photon flux. 
Resonance enhancements are observed in two distinct structures: One at the binding energy ($E_B$) of $4-8$ eV and another near $E_F$ at $E_B$ = $0-3$ eV. 
Additionally, the peak maximum of the structure near $E_F$ exhibits a shift in binding energy depending on $h\nu$. 
This behavior is likely due to multiplet effects where the photon energy with the maximum enhancement depends on the component of the final-state multiplet, which have been similarly reported in PrTi$_2$Al$_{20}$.\cite{Pr1220_RPES}

To clarify the origin of the resonance structures at $E_B$ = $4-8$ eV and near $E_F$, we plot a constant-initial-state (CIS) spectra at $E_B$ = 5.65 and 0.3 eV, labeled as CIS$_\alpha$ and CIS$_\beta$, in Fig.~\ref{RPES2Dmap1220}(b). 
The incident photon energy of the CIS$_\alpha$ peak is close to that of the absorption peak, indicating that the structure at $E_B$ = $4-8$ eV originates from the localized $4f^2$ final states. 
In contrast, the incident photon energy of the CIS$_\beta$ peak is clearly lower than that of the absorption peak, suggesting a higher number of the $4f$ electrons. 
Therefore, it is concluded that the structure near $E_F$ originates from the $4f^3$ final states. 

\begin{figure}[t] 
  \centering
       \includegraphics[keepaspectratio,width=80mm]{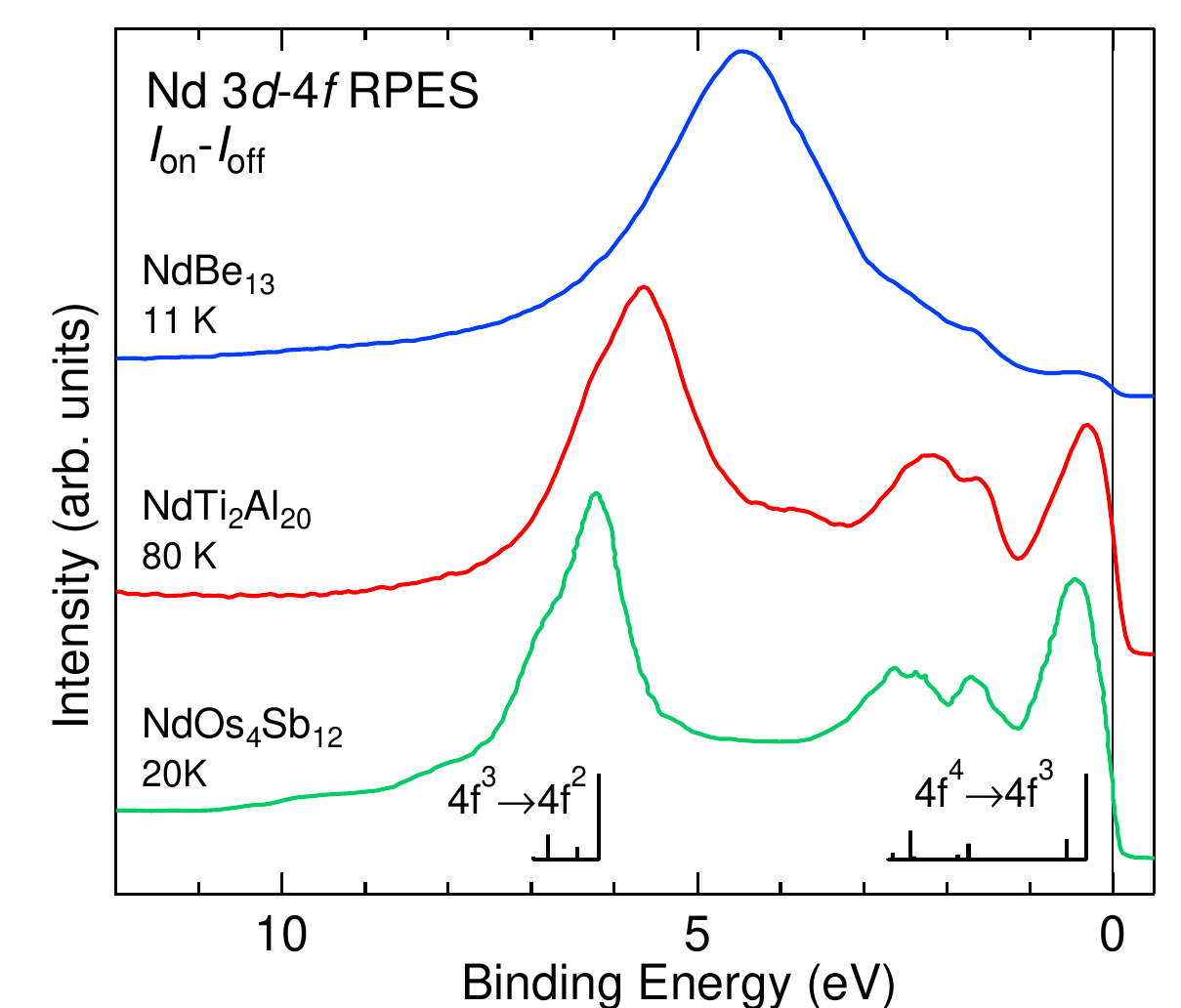}
 \caption
 {Resonance valence-band photoemission spectra obtained by subtracting the off-RPES spectra from the on-RPES spectra of NdTi$_2$Al$_{20}$ and NdBe$_{13}$ in Figs.~\ref{XASandRPES}(b,c). For comparison, the result of NdOs$_4$Sb$_{12}$ (Ref.\onlinecite{NdOsSb}) is also plotted. The vertical black lines indicate the $4f^2$ and $4f^3$ final-state multiplet components due to the 4f-derived photoemission processes.\cite{multiplet}} 
   \label{RPEScomp}
\end{figure}

To extract the resonance components in the Nd sites, we have subtracted the off-RPES spectrum from the on-RPES spectrum for NdTi$_2$Al$_{20}$ and NdBe$_{13}$ as shown in Fig.~\ref{RPEScomp}. 
For comparision, the spectrum of so far reported\cite{NdOsSb} NdOs$_4$Sb$_{12}$ is also plotted in the figure. 
For NdTi$_2$Al$_{20}$, structures are observed at $E_B$ = $4-8$ eV and from $E_F$ to $\sim$3 eV. 
The structure at $E_B$ = $4-8$ eV is similarly observed also for NdBe$_{13}$ and NdOs$_4$Sb$_{12}$, which is attributed from the $4f^3\to4f^2$ process, namely, the $4f^2$ final-state multiplets.  
The structure from $E_F$ to $\sim$3 eV is also seen for NdOs$_4$Sb$_{12}$ while this is not clearly observed for NdBe$_{13}$. 
This structure with several peaks and/or shoulders is well explained as the $4f^4\to4f^3$ process final-state multiplets shown at the bottom of Fig.~\ref{RPEScomp}, which has been simulated by the single-ion model with the $4f^4$ initial state.\cite{multiplet} 
Therefore, we conclude that this is attributed to the $4f^3$ final states arising from the $c$-$f$ hybridization. 

The resonance spectral enhancement near $E_F$ is also seen for NdBe$_{13}$,but the enhancement itself is relatively much weak. 
We have also compared the CIS spectra near $E_F$ with the XAS spectrum of NdBe$_{13}$, from which the fact that the incident photon energies of these peaks are mutually identical (see the APPENDIX). 
This tendency is qualitatively different from that for NdTi$_2$Al$_{20}$. 
Thus, it is concluded that the structure near $E_F$ for NdBe$_{13}$ is not due to the 4$f^3$ final-state component but originating from the Nd $5d$ components ejected through Auger process.\cite{Auger2013,Auger2015,Auger2023} 

\subsection{Discussion}
The center of gravity for the 4$f^2$ final-state components in the RPES spectra depends strongly on the material. 
The energy splitting of the gravities of the $4f^3$ and $4f^2$ final-state multiplets is slightly smaller for NdTi$_2$Al$_{20}$ than for NdOs$_4$Sb$_{12}$. 
The center of the gravity for the $4f^2$ multiplets is located at the shallowest $E_B$ for NdBe$_{13}$, of which the $4f^3$ final-state components have not been confirmed. 
As pointed out above, $E_B$ of the Nd $3d_{5/2}$ core-level main peak is shallower for NdBe$_{13}$ than for NdTi$_2$Al$_{20}$. 
In addition, the latter shows the clear $3d^94f^4\underline{c} $ final states at $\sim$974 eV which are not recognized for the former in the Nd $3d_{5/2}$ core-level PES spectra. 
When the energy differences in the bare Nd $3d$ and $4f$ levels with composition are relatively smaller than the energy scale of the $c$-$f$ hybridizations, which would be the case for the present systems,  
these results suggest that the binding energies of the main peaks due to the $4f^3$ initial states in both core-level and valence-band PES spectra reflect the degree of the $c$-$f$ hybridization effects. 
Namely, the $c$-$f$ hybridization has a essential role for the $4f$ electronic states of NdTi$_2$Al$_{20}$ while its hybridization effect is relatively weaker than that for NdOs$_4$Sb$_{12}$. 
On the other hand, the $4f$ states are well localized due to the clearly weaker $c$-$f$ hybridization for NdBe$_{13}$. 

It is clear that the $c$-$f$ hybridization effect is responsible for the two-channel Kondo effect. 
This effect has also been observed by the RPES of PrTi$_2$Al$_{20}$, where the two-channel Kondo (quadrupolar Kondo) effect has been discussed.\cite{Pr_Two} 
Furthermore, for PrIr$_2$Zn$_{20}$ with the same crystal structure as that for PrTi$_2$Al$_{20}$, linear dichroism (LD) in the Pr 3$d$ core-level HAXPES spectra has revealed that the CEF-split ground state symmetry is in the $\Gamma_3$ symmetry, supporting the two-channel Kondo (quadrupolar Kondo) scenario.\cite{PrLD1,PrLD2} 
In the case of the Nd systems, it is suggested that $\Gamma_6$ symmetry is required to induce the two-channel Kondo effect. 
Therefore, by determining the symmetry in NdTi$_2$Al$_{20}$ through the Nd 3$d$ HAXPES-LD, it will be possible to more precisely examine the potential for the two-channel Kondo effect in future.

\section{SUMMARY}
We have performed the HAXPES, SXPES and XAS on the two-channel Kondo effect candidate NdTi$_2$Al$_{20}$ and the weak $c$-$f$ hybridization compound NdBe$_{13}$ of which both show the AFM transitions at low temperatures. 
The spectroscopic evidence for the notable $c$-$f$ hybridization effects, which is necessary for the two-channel Kondo effect, is clearly shown for NdTi$_2$Al$_{20}$ while it is absent for NdBe$_{13}$. 
It is also found that the difference in the $4f$ hybridizations between the intermetallic Nd compounds and the insulating Nd-based oxides is reflected in the bulk 3d core-level spectra. 
While the $4f^2$ final-state multiplet structure is seen for both hybridized and $4f$-localized intermetallic Nd compounds, its center of gravity depends on the material reflecting the different $c$-$f$ hybridization effects.

\section*{ACKNOWLEDGMENTS}
We thank S. Nakajima and K. Kuga for supporting the experiments. 
The HAXPES experiments at SPring-8 BL19LXU were performed under the approval of RIKEN and JASRI (Proposal Nos. 20160034, 20220071, 20230087, 20240066, 2024A1398, and 2024B1511) whereas  the XAS and RPES experiments at SPring-8 BL17SU were performed under the approval of RIKEN (Proposals Nos. 20220027, 20230077 and 20240072) at SPring-8. 
The HAXPES experiments in Ritsumeikan University SR center shown in Fig.~\ref{Nd3dn} were performed under the approval of SR center (Proposals Nos. S23013, S23026). 
This work was financially supported by a Grant-in-Aid for Innovative Areas (JP19H05817, JP19H05818, JP20H05271, and JP22H04594), a Grant-in-Aid for Transformative Research (JP23H04867), a Grant-in-Aid for Scientific Research (JP20K20900, JP22K03527, JP24KJ1587, and JP24K03202) from JSPS and MEXT. 
G. Nozue was supported by the The University of Osaka fellowship program of Super Hierarchical Materials Science Program and by the JSPS Research Fellowships for Young Scientists.

\section*{APPENDIX: $\rm{\textbf{Valence-band}}$ RPES OF $\rm{\textbf{NdBe}}_{\textbf{13}}$}
\vspace{-2mm}
\begin{figure}[htbp] 
  \centering
       \includegraphics[keepaspectratio,width=80mm]{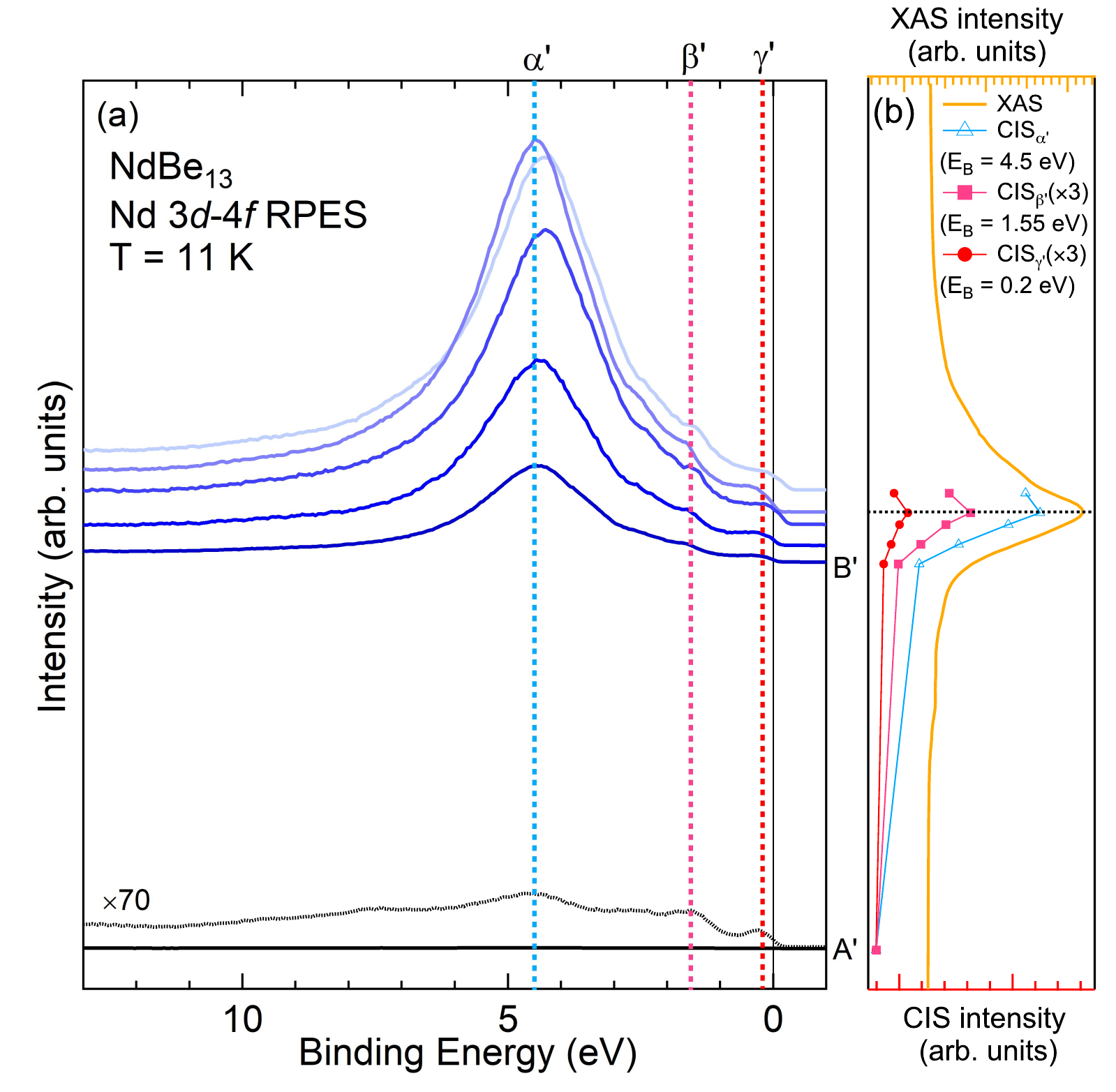}
 \caption
 {(a) Nd $3d$-$4f$ valence-band RPES spectra of NdBe$_{13}$ across the Nd $M_5$ edge at six different incident photon energies. The spectra A$'$ and B$'$ correspond to that in Fig.~\ref{Nd3dn}(c). 
(b) CIS spectra at $E_B$ = 4.5 (${\alpha'}$), 1.55 (${\beta'}$), and 0.2 eV (${\gamma'}$). 
For comparison, the experimental XAS spectrum (yellow) is also plotted. 
The black dotted line indicates the incident photon energy of the absorption peak.} 
   \label{RPESNdBe13}
\end{figure}

We have performed the RPES of NdBe$_{13}$ across the Nd $M_5$ edge at six different insident photon energies. Figure~\ref{RPESNdBe13}(a) shows the RPES spectra normalized by the photon flux where A$'$ and B$'$ labeled in the figure correspond to that in Fig.~\ref{Nd3dn}. 
We have compared the CIS spectra at $E_B$ = 4.5 ($\alpha'$), 1.55 ($\beta'$), and 0.2 eV ($\gamma'$) with the XAS spectrum as shown in Fig.~\ref{RPESNdBe13}(b). 
In the same manner as those of NdTi$_2$Al$_{20}$ in Fig.~\ref{RPES2Dmap1220}, CIS$_{\alpha'}$ originating from the localized $4f^2$ final states exhibits the peak at nearly the same incident energy as that of the XAS spectrum. 
On the other hand, the incident photon energy of the peaks in CIS$_{\beta'}$ and CIS$_{\gamma'}$, at $E_B$ of the structure around $E_F$, is also highly consistent with the peak of the XAS peak, suggesting that the structures near $E_F$ are not due to the $4f^3$ final-state components arising from the hybridization.

\pagebreak



\begin{thebibliography}{99}
\bibitem{Kondo1}
J. Kondo, Prog. Theor. Phys. \textbf{32}, 37 (1964). 
\bibitem{Kondo2}
K. Yosida, Phys. Rev. \textbf{147}, 223 (1966). 
\bibitem{Kondo_Yb}
M. Kasaya, F. Iga, K. Negishi, S. Nakai, and T. Kasuya, J. Magn. Magn. Mater. \textbf{31-34}, 437 (1983). 
\bibitem{Kondo_Ce}
N. Sato, A. Sumiyama, S. Kunii, H. Nagano, and T. Kasuya, J. Phys. Soc. Jpn. \textbf{54}, 1923 (1985). 
\bibitem{NdCo}
R. Yamamoto, R. J. Yamada, Y. Yamane, Y. Shimura, K. Umeo, T. Takabatake, and T. Onimaru,  Phys. Rev. B \textbf{104}, 155112 (2021).
\bibitem{Two_Cox}
D. L. Cox, Phys. Rev. Lett. \textbf{59}, 1240 (1987).
%
\bibitem{LLW}
K. R. Lea, M. J. M. Leask, and W. P. Wolf, J. Phys. Chem. Solids \textbf{23}, 1381 (1962).
\bibitem{Kondo_Nd}
T. Hotta, J. Phys. Soc. Jpn. \textbf{86}, 083704 (2017).



\bibitem{Nd1220_crystal}
T. Namiki, K. Nosaka, K. Tsuchida, Q. Lei, R. Kanamori, and K. Nishimura, J. Phys.: Conf. Ser. \textbf{683}, 012017 (2016).
%
\bibitem{Nd1220_Sugawara2020}
T. Yoshinaga, J. Kaneyoshi, E. Matsuoka, H. Kotegawa, H. Tou, A. Nakamura,
D. Aoki, H. Harima, and H. Sugawara, J. Phys.: Conf. Ser. \textbf{30}, 011116 (2020).
\bibitem{NaZn13_cry}
D. P. Shoemaker, R. E. Marsh, F. J. Ewing, and L. Pauling, Acta Cryst. \textbf{5}, 637 (1952).
%
\bibitem{NdBe13_crystal}
M.J. Besnus, P. Panissod, J.P. Kappler, G. Heinrich and A. Meyer, J. Magn. Magn. Mater. \textbf{31-34}, 227 (1983).
\bibitem{NdBe13_AFM}
E. Bucher, J. P. Maita, G. W. Hull, R. C. Fulton, and A. S. Cooper, Phys. Rev. B \textbf{11}, 440 (1975).
%
\bibitem{HidakaUnpub}
H. Hidaka, T. Yanagisawa, and H. Amitsuka, unpublished.
\bibitem{BL17_1}
H. Ohashi, Y. Senba, H. Kishimoto, T. Miura, E. Ishiguro, T. Takeuchi, M. Oura, K. Shirasawa, T. Tanaka, M. Takeuchi, K. Takeshita, S. Goto, S. Takahashi, H. Aoyagi, M. Sano, Y. Furukawa, T. Ohata, T. Matsushita, Y. Ishizawa, S. Taniguchi, Y. Asano, Y. Harada, T. Tokushima, K. Horiba, H. Kitamura, T. Ishikawa, and S. Shin, AIP Conf. Proc. \textbf{879}, 523 (2007). 
%
\bibitem{BL17_2}
Y. Senba, H. Ohashi, H. Kishimoto, T. Miura, S. Goto, S. Shin, T. Shintake, and  T. Ishikawa, AIP Conf. Proc. \textbf{879}, 718 (2007). 
%
\bibitem{BL17_3}
T. Tanaka, T. Seike, A. Kagamihara, H. Aoyagi, T. Kai, M. Sano, S. Takahashi, and M. Oura, J. Synchrotron Rad.  \textbf{30}, 301 (2023). 
\bibitem{BL19}
H. Fujiwara, S. Naimen, A. Higashiya, Y. Kanai, H. Yomosa, K. Yamagami, T. Kiss, T. Kadono, S. Imada, A. Yamasaki, K. Takase, S. Otsuka, T. Shimizu, S. Shingubara, S. Suga, M. Yabashi, K. Tamasaku, T. Ishikawa, and A. Sekiyama, J. Synchrotron Radiat. \textbf{23}, 735 (2016).
\bibitem{Ritsu}
H. Fujiwara, A. Enomoto, M. Sakaguchi, S. Nakajima, A. Sekiyama, A. Irizawa, and S. Imada, Mem. SR Center Ritsumeikan Univ. \textbf{26}, 16 (2024).
\bibitem{NdOsSb}
S. Imada, H. Higashimichi, A. Yamasaki, M. Yano, T. Muro, A. Sekiyama, S. Suga, H. Sugawara, D. Kikuchi and H. Sato, Phys. Rev. B. \textbf{76}, 153106 (2007).
\bibitem{G_NdOsSb}
P.-C. Ho, W. M. Yuhasz, N. P. Butch, N. A. Frederick, T. A. Sayles, J. R. Jeffries, M. B. Maple, J. B. Betts, A. H. Lacerda, P. Rogl, and G. Giester, Phys. Rev. B \textbf{72}, 094410 (2005).
\bibitem{G_NdFeP}
D. Aoki, Y. Haga, Y. Homma, H. Sakai, S. Ikeda, Y. Shiokawa, E. Yamamoto, A. Nakamura and Y. Onuki, J. Phys.  Soc.  Jpn. \textbf{75}, 073703 (2006).
%
\bibitem{PrLD2}
S. Hamamoto, Y. Kanai, S. Fujioka, Y. Nakatani, H. Fujiwara, K. Kuga, T. Kiss,
A. Higashiya, A. Yamasaki, S. Imada, A. Tanaka, K. Tamasaku, M. Yabashi, T. Ishikawa,
H. Hidaka, T. Yanagisawa, H. Amitsuka, K. T. Matsumoto, T. Onimaru, T. Takabatake, and 
A. Sekiyama, J. Electron. Spectrosc. Relat. Phenom. \textbf{238}, 146885 (2020).
\bibitem{Suzuki}
C. Suzuki, J. Kawai, M. Takahashi, A.-H. Vlaicu, H. Adachi, T. Mukoyama, Chem. 
Phys. \textbf{253}, 27 (2000).
\bibitem{Horio}
M. Horio, K. Hauser, Y. Sassa, Z. Mingazheva, D. Sutter, K. Kramer, A. Cook, E. Nocerino, O. K. Forslund, O. Tjernberg, M. Kobayashi, A. Chikina, N.B.M. Schr{\"o}ter, J. A. Krieger, T. Schmitt, V. N. Strocov, S. Pyon, T. Takayama, H. Takagi, O. J. Lipscombe, S.M. Hayden, M. Ishikado, H. Eisaki, T. Neupert, M. M{\aa}nsson, C. E. Matt, and J. Chang, Phys. Rev. Lett. \textbf{121}, 077004 (2018).
%
\bibitem{ASNSMO99}
A. Sekiyama, S. Suga, M. Fujikawa, S. Imada, T. Iwasaki, K. Matsuda, T. Matsushita, K. V. Kaznacheyev, A. Fujimori, H. Kuwahara, and Y. Tokura, Phys. Rev. B \textbf{59}, 15528 (1999).
\bibitem{Pr1220_RPES}
M. Matsunami, M. Taguchi, A. Chainani, R. Eguchi, M. Oura, A. Sakai, S. Nakatsuji and S. Shin, Phys. Rev. B \textbf{84}, 193101 (2011).
\bibitem{multiplet}
F. Gerken, J. Phys. F: Met. Phys. \textbf{13}, 703 (1983).
\bibitem{Auger2013}
A. Yasui, Y. Saitoh, S.-i. Fujimori, I. Kawasaki, T. Okane, Y. Takeda, G. Lapertot, G. Knebel, T. D. Matsuda, Y. Haga, and H. Yamagami, Phys. Rev. B \textbf{87}, 075131 (2013).
\bibitem{Auger2015}
M.Y. Kimura, K. Fukushima, H. Takeuchi, S. Ikeda, H. Sugiyama, Y. Tomida, G. Kuwahara, H. Fujiwara, T. Kiss, A. Yasui, I. Kawasaki, H. Yamagami, Y. Saitoh, T. Muro, T. Ebihara, and A. Sekiyama, J. Phys. Conf. Ser. \textbf{592}, 012003 (2015).
\bibitem{Auger2023}
Y. Nakatani, H. Fujiwara, H. Aratani, T. Mori, S. Tachibana, T. Yamaguchi, T. Kiss, A. Yamasaki, A. Yasui, H. Yamagami, A. Tsuruta, J. Miyawaki, T. Ebihara, Y. Saitoh, A. Sekiyama, J. Electron. Spectrosc. Relat. Phenom. \textbf{220}, 50 (2017).
\bibitem{Pr_Two}
A. Sakai and S. Nakatsuji, J. Phys. Soc. Jpn. \textbf{80}, 063701 (2011).
%
\bibitem{PrLD1}
S. Hamamoto, S. Fujioka, Y. Kanai, K. Yamagami, Y. Nakatani, K. Nakagawa, H. Fujiwara, T. Kiss,
A. Higashiya, A. Yamasaki, T. Kadono, S. Imada, A. Tanaka, K. Tamasaku, M. Yabashi, T. Ishikawa,
K. T. Matsumoto, T. Onimaru, T. Takabatake, and  A. Sekiyama, 
J. Phys. Soc. Jpn. \textbf{86}, 123703 (2017).


\end{thebibliography}
\end{document}